\documentstyle[twoside]{article}

\catcode`\@=11
\long\def\@makefntext#1{
\protect\noindent \hbox to 3.2pt {\hskip-.9pt  
$^{{\eightrm\@thefnmark}}$\hfil}#1\hfill}		%CAN BE USED 

\def\thefootnote{\fnsymbol{footnote}}
\def\@makefnmark{\hbox to 0pt{$^{\@thefnmark}$\hss}}	%ORIGINAL 
	
\def\ps@myheadings{\let\@mkboth\@gobbletwo
\def\@oddhead{\hbox{}
\rightmark\hfil\eightrm\thepage}   
\def\@oddfoot{}\def\@evenhead{\eightrm\thepage\hfil
\leftmark\hbox{}}\def\@evenfoot{}
\def\sectionmark##1{}\def\subsectionmark##1{}}

\oddsidemargin=\evensidemargin
\renewcommand{\thefootnote}{\fnsymbol{footnote}}

\newcounter{sectionc}\newcounter{subsectionc}\newcounter{subsubsectionc}
\renewcommand{\section}[1] {\vspace{12pt}\addtocounter{sectionc}{1} 
\setcounter{subsectionc}{0}\setcounter{subsubsectionc}{0}\noindent 
	{\tenbf\thesectionc. #1}\par\vspace{5pt}}
\renewcommand{\subsection}[1] {\vspace{12pt}\addtocounter{subsectionc}{1} 
	\setcounter{subsubsectionc}{0}\noindent 
	{\bf\thesectionc.\thesubsectionc. {\kern1pt \bfit #1}}\par\vspace{5pt}}
\renewcommand{\subsubsection}[1] {\vspace{12pt}\addtocounter{subsubsectionc}{1}
	\noindent{\tenrm\thesectionc.\thesubsectionc.\thesubsubsectionc.
	{\kern1pt \tenit #1}}\par\vspace{5pt}}
\newcommand{\nonumsection}[1] {\vspace{12pt}\noindent{\tenbf #1}
	\par\vspace{5pt}}

\newcounter{appendixc}
\newcounter{subappendixc}[appendixc]
\newcounter{subsubappendixc}[subappendixc]
\renewcommand{\thesubappendixc}{\Alph{appendixc}.\arabic{subappendixc}}
\renewcommand{\thesubsubappendixc}
	{\Alph{appendixc}.\arabic{subappendixc}.\arabic{subsubappendixc}}

\renewcommand{\appendix}[1] {\vspace{12pt}
        \refstepcounter{appendixc}
        \setcounter{figure}{0}
        \setcounter{table}{0}
        \setcounter{lemma}{0}
        \setcounter{theorem}{0}
        \setcounter{corollary}{0}
        \setcounter{definition}{0}
        \setcounter{equation}{0}
        \renewcommand{\thefigure}{\Alph{appendixc}.\arabic{figure}}
        \renewcommand{\thetable}{\Alph{appendixc}.\arabic{table}}
        \renewcommand{\theappendixc}{\Alph{appendixc}}
        \renewcommand{\thelemma}{\Alph{appendixc}.\arabic{lemma}}
        \renewcommand{\thetheorem}{\Alph{appendixc}.\arabic{theorem}}
        \renewcommand{\thedefinition}{\Alph{appendixc}.\arabic{definition}}
        \renewcommand{\thecorollary}{\Alph{appendixc}.\arabic{corollary}}
        \renewcommand{\theequation}{\Alph{appendixc}.\arabic{equation}}
        \noindent{\tenbf Appendix \theappendixc #1}\par\vspace{5pt}}
\newcommand{\subappendix}[1] {\vspace{12pt}
        \refstepcounter{subappendixc}
        \noindent{\bf Appendix \thesubappendixc. {\kern1pt \bfit #1}}
	\par\vspace{5pt}}
\newcommand{\subsubappendix}[1] {\vspace{12pt}
        \refstepcounter{subsubappendixc}
        \noindent{\rm Appendix \thesubsubappendixc. {\kern1pt \tenit #1}}
	\par\vspace{5pt}}

\newcommand{\textlineskip}{\baselineskip=13pt}
\newcommand{\smalllineskip}{\baselineskip=10pt}

\def\eightcirc{
\begin{picture}(0,0)
\put(4.4,1.8){\circle{6.5}}
\end{picture}}
\def\eightcopyright{\eightcirc\kern2.7pt\hbox{\eightrm c}} 

\newcommand{\copyrightheading}[1]
        {\vspace*{-2.5cm}\smalllineskip{\flushleft
        {\footnotesize International Journal of Modern Physics A, #1}\\
        {\footnotesize $\eightcopyright$\, World Scientific Publishing
         Company}\\
         }}

\def\abstracts#1#2#3{{
	\centering{\begin{minipage}{4.5in}\baselineskip=10pt\footnotesize
	\parindent=0pt #1\par 
	\parindent=15pt #2\par
	\parindent=15pt #3
	\end{minipage}}\par}} 

\renewenvironment{thebibliography}[1]
	{\frenchspacing
	 \ninerm\baselineskip=11pt
	 \begin{list}{\arabic{enumi}.}
	{\usecounter{enumi}\setlength{\parsep}{0pt}
	 \setlength{\leftmargin 12.7pt}{\rightmargin 0pt} %FOR 1--9 ITEMS
	 \setlength{\itemsep}{0pt} \settowidth
	{\labelwidth}{#1.}\sloppy}}{\end{list}}

\newcounter{itemlistc}
\newcounter{romanlistc}
\newcounter{alphlistc}
\newcounter{arabiclistc}

\newcommand{\fcaption}[1]{
        \refstepcounter{figure}
        \setbox\@tempboxa = \hbox{\footnotesize Fig.~\thefigure. #1}
        \ifdim \wd\@tempboxa > 5in
           {\begin{center}
        \parbox{5in}{\footnotesize\smalllineskip Fig.~\thefigure. #1}
            \end{center}}
        \else
             {\begin{center}
             {\footnotesize Fig.~\thefigure. #1}
              \end{center}}
        \fi}

\newcommand{\tcaption}[1]{
        \refstepcounter{table}
        \setbox\@tempboxa = \hbox{\footnotesize Table~\thetable. #1}
        \ifdim \wd\@tempboxa > 5in
           {\begin{center}
        \parbox{5in}{\footnotesize\smalllineskip Table~\thetable. #1}
            \end{center}}
        \else
             {\begin{center}
             {\footnotesize Table~\thetable. #1}
              \end{center}}
        \fi}

\def\@citex[#1]#2{\if@filesw\immediate\write\@auxout
	{\string\citation{#2}}\fi
\def\@citea{}\@cite{\@for\@citeb:=#2\do
	{\@citea\def\@citea{,}\@ifundefined
	{b@\@citeb}{{\bf ?}\@warning
	{Citation `\@citeb' on page \thepage \space undefined}}
	{\csname b@\@citeb\endcsname}}}{#1}}

\newif\if@cghi
\def\cite{\@cghitrue\@ifnextchar [{\@tempswatrue
	\@citex}{\@tempswafalse\@citex[]}}
\def\citelow{\@cghifalse\@ifnextchar [{\@tempswatrue
	\@citex}{\@tempswafalse\@citex[]}}
\def\@cite#1#2{{$\null^{#1}$\if@tempswa\typeout
	{IJCGA warning: optional citation argument 
	ignored: `#2'} \fi}}

\def\pmb#1{\setbox0=\hbox{#1}
	\kern-.025em\copy0\kern-\wd0
	\kern.05em\copy0\kern-\wd0
	\kern-.025em\raise.0433em\box0}

\def\fnt#1#2{\footnotetext{\kern-.3em
	{$^{\mbox{\scriptsize #1}}$}{#2}}}

\def\fpage#1{\begingroup
\voffset=.3in
\thispagestyle{empty}\begin{table}[b]\centerline{\footnotesize #1}
	\end{table}\endgroup}

\def\runninghead#1#2{\pagestyle{myheadings}
\markboth{{\protect\footnotesize\it{\quad #1}}\hfill}
{\hfill{\protect\footnotesize\it{#2\quad}}}}
\font\tenrm=cmr10
\font\tenit=cmti10 
\font\tenbf=cmbx10
\font\bfit=cmbxti10 at 10pt
\font\ninerm=cmr9

\font\eightrm=cmr8

\def\qed{\hbox{${\vcenter{\vbox{			%HOLLOW SQUARE
   \hrule height 0.4pt\hbox{\vrule width 0.4pt height 6pt
   \kern5pt\vrule width 0.4pt}\hrule height 0.4pt}}}$}}

\renewcommand{\thefootnote}{\fnsymbol{footnote}}	%USE SYMBOLIC FOOTNOTE

\begin{document}

\normalsize\textlineskip
\thispagestyle{empty}
\setcounter{page}{1}

\copyrightheading{}                     %{Vol. 0, No. 0 (1993) 000--000}

\vskip 0.2cm

\flushright{\large SMU-HEP-00-17}

\flushleft

\runninghead{Yongsheng Gao, Southern Methodist Univ., Talk at DPF2000, 
             Columbus, OHIO} 
  {Yongsheng Gao, Southern Methodist Univ., Talk at DPF2000, Columbus, OHIO}

\vspace*{0.88truein}

\fpage{1}
\centerline{\bf }
\vspace*{0.035truein}
\centerline{\bf Results on $B$ to Charmonia Decays from CLEO}
\vspace*{0.37truein}
\centerline{\footnotesize YONGSHENG GAO\footnote{Physics Department,
             Southern Methodist University, Dallas, TX 75275, USA}}
\vspace*{0.015truein}
\centerline{\footnotesize\it Physics Department, Southern Methodist
                             University}
\baselineskip=10pt
\centerline{\footnotesize\it Dallas, TX 75275, USA
            \footnote{Email: gao@mail.physics.smu.edu}}
\vspace*{10pt}
%\vspace*{0.225truein}
%\publisher{(received date)}{(revised date)}

\vspace*{0.21truein}
\abstracts{We review some recent results on $B$ to charmonia decays 
from CLEO based on $9.7 \times 10^6$ $\mbox{$B\bar{B}$}$ pairs 
collected with the CLEO detector. 
These include measurements of
$B^{0} \to J/\psi K^{0}_{s}$, $\chi_{c1} K^{0}_{s}$ and
$J/\psi \pi^{0}$ branching fractions, search for direct $CP$ violation
in $B^{\pm} \to J/\psi K^{\pm}$ and $\psi(2S) K^{\pm}$,
observation of $B \to \eta_{c} K$,
study of $\chi_{c1}$, $\chi_{c2}$ production and measurement of
$f_{+-}/f_{00} \equiv {\cal B}[\Upsilon(4S) 
     \to B^+ B^-]/{\cal B}[\Upsilon(4S) \to B^0 \overline B^0]$.}{}{}

%\textlineskip			%) USE THIS MEASUREMENT WHEN THERE IS
%\vspace*{12pt}			%) NO SECTION HEADING
 
\vspace*{1pt}\textlineskip	%) USE THIS MEASUREMENT WHEN THERE IS
\vskip 0.3cm
\begin{center}
Talk given at DPF2000, Columbus, Ohio, USA, August 9 $-$ 12, 2000. 
\end{center}

\textheight=7.8truein
\setcounter{footnote}{0}
\renewcommand{\thefootnote}{\alph{footnote}}

\section{CESR and CLEO}
\noindent
The Cornell Electron Storage Ring (CESR) is a symmetric $e^+ e^-$ 
collider operating at the $\Upsilon$(4S) resonance.  
The CLEO II and II.V detector configurations are described in detail 
elsewhere~\cite{cleoii,cleoiiv}.
The integrated luminosity is 9.2 fb$^{-1}$ at the $\Upsilon$(4S) 
resonance, which corresponds to about 9.7 $\times$ 10$^6$
$\mbox{$B\bar{B}$}$ pairs, and 4.6 fb$^{-1}$ at energies just below 
the $\mbox{$B\bar{B}$}$ threshold.
The results reviewed in this paper are based on this full data sample.

\section{$B^{0}$ Decays for sin2$\beta$ measurement}
\noindent
The origin of $CP$ violation is one of the most important problems
of experimental high energy physics.
The study of $B$ mesons has been attracting extensive world wide
attention because it will allow for a decisive test of the
quark-mixing sector in the Standard Model (SM).
It is very important to test whether the SM provides the 
correct description of $CP$ violation, in order to search for 
new physics beyond the SM. 
That is the major motivation for building the current generation of 
$B$ factories, BABAR, BELLE and CLEO III. 
To check the SM predictions on $CP$ violation in order to
search for new physics beyond the SM, it is particularly
important to measure the three sides and angles 
($\alpha$ $\equiv$ $arg$ [$- \frac{V_{td}V_{tb}^{*}}{V_{ud}V_{ub}^{*}}$],
$\beta$  $\equiv$ $arg$ [$- \frac{V_{cd}V_{cb}^{*}}{V_{td}V_{tb}^{*}}$], 
and
$\gamma$ $\equiv$ $arg$ [$- \frac{V_{ud}V_{ub}^{*}}{V_{cd}V_{cb}^{*}}$])
of the CKM unitary triangle as precise as possible.
Among these three CKM angles, only the angle $\beta$ (sin2$\beta$)
is expected to be cleanly measured in the near future using the gold 
plated $B^{0} \to J/ \psi K^0_s$ decay mode.

\vskip 0.4cm

\noindent
The first goal of these first-generation $B$ factories is to observe
$CP$ violation in the $B$ meson system which is predicted in the SM.
Strong evidence of $CP$ violation in the $B$ system
has been found through the Sin2$\beta$ measurement from 
CDF~\cite{cdf2beta}. Better measurements from BABAR and BELLE 
were shown this summer with lower central value than 
expected~\cite{babar2beta,belle2beta}. Besides the golden decay mode
$B^{0} \to J/\psi K^{0}_{S}$ where $K^{0}_{S} \to \pi^{+}\pi^{-}$
in the sin2$\beta$ measurement,
similar decay modes 
             $B^{0} \to J/\psi K^{0}_{S}$ where  
             $K^{0}_{S} \to \pi^{0}\pi^{0}$,
             $B^{0} \to \chi_{c1} K^{0}_{S}$,
             $B^{0} \to J/\psi \pi^{0}$ etc can also be used. 
These decay modes will add about 15\% more statistics to the
sin2$\beta$ measurement. This is very important when the
sin2$\beta$ measurements are dominated by statistics errors.
Furthermore, the tree-penguin interference in $B^{0} \to J/\psi \pi^{0}$
may allow the removal of the $\beta$ and $\beta$ + $\pi$ 
ambiguity~\cite{GrossmanQuinn}.

\subsection{CLEO results on $B^{0} \to J/\psi K^{0}_{S}$,
                $\chi_{c1} K^{0}$ and $J/\psi \pi^{0}$}

\noindent
Clean and significant signals in
$B^{0} \to J/\psi K^{0}_{S}(\pi^{0}\pi^{0})$, $\chi_{c1} K^{0}$ and 
$J/\psi \pi^{0}$ have been observed with very little backgrounds.
The main results are summarized in the following table.
The details of the analyses to obtain these results can be found 
in~\cite{cleocharmonium}.

\noindent

\begin{table}[htb]
\center
\begin{tabular}{rlcccl}
  & Decay mode          & Signal    & Background  &   
  Efficiency $(\%)$ & ${\cal B}$ $(\times10^{-4})$   \\ \hline
 & $B^0 \to J/\psi \, K^0$      &          &          &
       &  $9.5\pm0.8\pm0.6$   \\
Update  &  ~~~$K^0_S \to \pi^+ \pi^-$      &   142    & $0.3\pm
0.2$ &
$37.0\pm2.3$  & $9.8 \pm 0.8 \pm 0.7$                 \\
New     & ~~~$K^0_S \to \pi^0 \pi^0$       &   ~22    &  $1.1\pm0.3$
  &
$13.9\pm1.1$    & $8.4^{+2.1}_{-1.9} \pm 0.7$                 \\
New & $B^0 \to \chi_{c1}\,  K^0$           &    ~9    & $0.
9\pm0.3$  &
$19.2\pm1.3$   & $3.9^{+1.9}_{-1.3}\pm 0.4$                 \\
New  & $B^0 \to J/\psi \, \pi^0$           &   ~10    &
 $1.0\pm0.5$  &
$31.4\pm2.2$   & $0.25^{+0.11}_{-0.09}\pm 0.02$   \\
\end{tabular}
\end{table}

\section{Search for direct $CP$ violation in 
                $B^{\pm} \to \psi^{(\prime)} K^{\pm}$}
\noindent
Interfering amplitudes with different $CP$-even (strong and 
electromagnetic) and $CP$-odd (weak) phases are the necessary
ingredients for direct $CP$-violation. In the case of
$B^{\pm} \to \psi^{(\prime)} K^{\pm}$ decays, we do have
interfering tree and penguin diagrams which could have a significant
relative strong phase. However, the relative weak phase is very
small. As a result, the asymmetry in $B^{\pm} \to \psi^{(\prime)} K^{\pm}$ 
in SM is well below the 4\% expected precison of the CLEO measurement.
On the other hand, new physics beyond the SM can introduce large
direct $CP$ violation in $B^{\pm} \to \psi^{(\prime)} K^{\pm}$.
For instance, A $CP$ asymmetry of about 10\% in 
$B^{\pm} \to \psi^{(\prime)} K^{\pm}$ decays is 
possible in certain Two-Higgs doublet model~\cite{SoniWu}

\vskip 0.4cm

\noindent
The $CP$ asymmetry is defined as follows:

\vskip -0.4cm

\begin{eqnarray}
\nonumber {\cal A}_{CP} \equiv \frac{ {\cal B}(B^- \to \psi^{(\prime)} \, 
 K^-)- {\cal  B}(B^+ \to \psi^{(\prime)}  \, K^+)} {{\cal B}(B^-  \to  
 \psi^{(\prime)} \, K^-)+ {\cal B}(B^+ \to \psi^{(\prime)} \, K^+)} = 
 \frac{b- \overline b}{b+ \overline b}  
\end{eqnarray}

\noindent
About 534 $B^{\pm} \to J/\psi \, K^{\pm}$
signal events and 120 $B^{\pm} \to \psi(2S) \, K^{\pm}$ signal
events are observed. No sign of direct $CP$ violation is observed.
The results are summarized in the following table.
The details of the analysis to obtain these results can be found 
in~\cite{cleocpviolation}.

\begin{table}[htbp]
\begin{center}
\begin{tabular}{l c c c c c } 
 Mode   &   $N(B^{\pm})$ 
        &   $N(B^{-})$ 
        &   $N(B^{+})$ 
        &   $\frac{N(B^{-})-N(B^{+})}{N(B^{-})+N(B^{+})}$ 
        &   $ {\cal A}_{CP}$ 
\\ \hline 
 $B^{\pm} \to J/\psi \, K^{\pm}$ 
       &   534 
       &   271 
       &   263 
       &   $(1.5\pm 4.3)\%$ 
       &   $(1.8\pm 4.3\pm 0.4)\%$ 
\\ 
 $B^{\pm} \to \psi(2S) \, K^{\pm}$ 
       &   120 
       &   61 
       &   59 
       &   $(1.7\pm 9.1)\%$ 
       &   $(2.0\pm 9.1\pm 1.0)\%$ \\ 
\end{tabular}
\end{center}
\end{table}

\section{Observation of $B \to \eta_{c} K$}
\noindent
Unexpected large $B \to \eta^{,} X$ branching fractions were observed
at CLEO~\cite{cleoetaprime}. Among several theoretical explanations,
a substantial intrinsic charm component in the $\eta^{,}$ has been 
proposed. If this is the case, the $\eta^{,}$ can be produced by
the axial part of the $b \to c \bar{c} s(d)$ process, which also 
produces $\eta_{c}$. Exclusive $B$ decays to charmonium states are
also of theoretical interest as a testing ground for the QCD 
calculations of quark dynamics and factorization. In the absence of
enhancing mechanisms, the $B$ decay rate to $\eta_{c} X$ is expected
to be comparable to that for the $B$ decays to $J/\psi X$.
The color-singlet production of $\chi_{c0}$ in $B$ decays vanishes 
in the factorization approximation as a consequence of spin-parity
conservation. However, the color-octet mechanism allows for the
production of the $\chi_{c0}$ P-wave 0$^{++}$ state via the emission
of a soft gluon.

\vskip 0.4cm

\noindent
By performing maximum likelihood analysis, CLEO has observed the
decay $B \to \eta_{c} K$ in both charged and neutral modes with 
branching fractions similar to those for $B \to J/\psi K$.
The channel $B^{0} \to \eta_{c} K^{0}$ can be used to extract the
value of sin(2$\beta$) via future time-dependent asymmetry 
measurement. Upper limits on $B \to \chi_{c0} K$ decays are set
that restricts possible enhancement of the $\chi_{c0}$ production
due to the color-octet mechanism.
The details of the analyses can be found in~\cite{cleoetac}. 
The branching fractions are measured to be:

\begin{center}
${\cal B}(B^+ \to  \eta_c \, K^+)=$ 
                $( 0.69^{+0.26}_{-0.21} \pm 0.08 \pm 0.20 )\times 10^{-3}$

${\cal B}(B^0 \to  \eta_c \, K^0)=$ 
                $( 1.09^{+0.55}_{-0.42} \pm 0.12 \pm 0.31 )\times 
                10^{-3}$
\end{center}

\section{Study of $\chi_{c1}$ and $\chi_{c2}$ production in $B$ decays}
\noindent
Recent measurements of charmonium production at Tevatron~\cite{tevatron} 
have brought surprises that resulted in better calculations of the 
inclusive charmonium production. Inclusive $B$ decays to charmonium provides
another test ground where theoretical predictions can be confronted 
with experimental data. A measurement of the $\chi_{c2}$-to-$\chi_{c1}$
production ratio in $B$ decays provides an especially clean test
of charmonium production models. The $B \to \chi_{c2} X$ decay is
forbidden at leading order in $\alpha_{s}$ in the color-singlet
model~\cite{colorsinglet}, while this production ratio should be 
5:3 if the color-octet
mechanism dominates in  $B \to \chi_{cJ} X$~\cite{coloroctet}.

\vskip 0.4cm

\noindent
Significant $B \to \chi_{c1} X$ signals have been observed, while
the signal yield for $B \to \chi_{c2} X$ is consistent with zero.
The results are shown in the following table.
The details of the analysis to obtain these results can be found 
in~\cite{cleochic1}.

\begin{table}[hhh]
\begin{tabular}{lcc}
  Branching Ratio          &   Measured Value    
                           &   95\% C.L. Upper Limit        \\ \hline
 ${\cal B}(B \to \chi_{c1} X$  
                           &   (4.14$\pm$0.31$\pm$0.40)$\times 10^{-3}$
                           &                                \\
 ${\cal B}(B \to \chi_{c1}[direct] X$)  
                           &   (3.83$\pm$0.31$\pm$0.40)$\times 10^{-3}$
                           &                                \\ \hline
 ${\cal B}(B \to \chi_{c2} X$)  
                           &   (0.98$\pm$0.48$\pm$0.15)$\times 10^{-3}$
                           &   $<$ 2.0 $\times 10^{-3}$     \\
 ${\cal B}(B \to \chi_{c2}[direct] X$)  
                           &   (0.71$\pm$0.48$\pm$0.16)$\times 10^{-3}$
                           &   $<$ 1.7 $\times 10^{-3}$     \\ \hline
 $\frac{{\cal B}(B \to \chi_{c2}[direct] X)}
            {{\cal B}(B \to \chi_{c1}[direct] X)}$
                           &   (0.18$\pm$0.13$\pm$0.04)
                           &   $<$ 0.44                     \\ \hline

\end{tabular}
\end{table}

\section{Measurement of the Relative Branching Fraction of 
         $\Upsilon(4S)$ to Charged and Neutral B-Meson Pairs}
\noindent
All the branching ratio measurements from the $e^{+}e^{-}$
colliders operating at the $\Upsilon$(4S) assumes equal production
of charged and neutral $B$ mesons.
The best previous measurement of the admixture ratio of charged
to neutral $B$ meson production at the $\Upsilon$(4S) yields
a value accurate only to about 15\%~\cite{cleoold}. 
Better measurement of this ratio will result in better branching 
ratio measurements of all the charged and neutral $B$ decays at 
the  $\Upsilon$(4S) which dominate all the $B$ meson decays.

\vskip 0.4cm

\noindent
The relative branching fraction of $\Upsilon$(4S) to charged
and neutral $B$ mesons is defined as:
\begin{center}
$\frac{\,f_{+-}}{f_{00}}\equiv\frac{\,{\cal B}[\Upsilon(4S)\to B^+ B^-]}{
{\cal B}[\Upsilon(4S)\to B^0 \overline B^0]}$
\end{center}
This ratio is measured to be:
\begin{center}
$\frac{f_{+-}}{f_{00}}$=$1.04\pm0.07\pm0.04$ 
\end{center}
\noindent
The details of the analysis can be found in~\cite{cleof00}. 

\vskip 0.4cm

\nonumsection{Acknowledgments}
\noindent
I would like to thank Alexey Ershov, David Jaffe and my other CLEO 
colleagues for the helpful discussions in preparing for this talk.


\vskip 0.5cm

\begin{thebibliography}{000}
\bibitem{cleoii}  CLEO Collaboration, Y. Kubota {\it et al.}, Nucl. Instrum.
                  Methods  A {\bf 320}, 66 (1992).

\bibitem{cleoiiv} T.~Hill, Nucl. Instrum.
                  Methods  A {\bf 418}, 32 (1998).



% CDF sin2beta measurement
\bibitem{cdf2beta}T. Affolder {\it et al.}
         (CDF Collaboration), Phys. Rev. {\bf D61}, 072005 (2000).


% BaBar sin2beta measurement
\bibitem{babar2beta}B. Aubert {\it et al.}
         (BABAR Collaboration), hep-ex/0008048.


% Belle sin2beta measurement
\bibitem{belle2beta}Hiroaki Aihara, (BELLE Collaboration), 
        {\it talk at the XXXth International Conference on High
             Energy Physics, July 27 $-$ August 2, 2000, Osaka, Japan}.


\bibitem{GrossmanQuinn}
Y. Grossman and H. R. Quinn, Phys.Rev. {\bf D56} 7259 (1997).


\bibitem{cleocharmonium}
P. Avery {\it et al.} (CLEO Collaboration),
 Phys. Rev. {\bf D62} 051101 (2000).


\bibitem{SoniWu}
G.H. Wu, A. Soni,
Phys. Rev. {\bf D62} 056005 (2000).


\bibitem{cleocpviolation}
G. Bonvicini {\it et al.} (CLEO Collaboration),
 Phys. Rev. Lett. {\bf 84} 5940 (2000).


\bibitem{cleoetaprime}
T. E. Browder {\it et al.} (CLEO Collaboration),
Phys. Rev. Lett. {\bf 81} 1786 (1998);
B. Behrens {\it et al.} (CLEO Collaboration),
Phys. Rev. Lett. {\bf 80} 3710 (1998).


\bibitem{cleoetac}
K.W. Edwards {\it et al.} (CLEO Collaboration),
hep-ex/0007012, CLNS 00/1680, CLEO 00-12.


\bibitem{tevatron}
F. Abe {\it et al.} (CDF Collaboration),
Phys. Rev. Lett. {\bf 69} 3704 (1992);
F. Abe {\it et al.} (CDF Collaboration),
Phys. Rev. Lett. {\bf 79} 572 (1997);
F. Abe {\it et al.} (CDF Collaboration),
Phys. Rev. Lett. {\bf 79} 578 (1997);
S. Abachi {\it et al.} (D0 Collaboration),
Phys. Lett. {\bf B370} 239 (1996).


\bibitem{colorsinglet}
J. H. Kuhn, S. Nussinov and R. Ruckl,
Z. Phys. {\bf C5} 117 (1980); 
J. H. Kuhn and R. Ruckl,
Phys. Lett. {\bf B135} 477 (1984); 
J. H. Kuhn and R. Ruckl,
Phys. Lett. {\bf B258} 499 (1991).


\bibitem{coloroctet}
G. A. Schuler, Eur. Phys. J. {\bf C8} 273 (1999).


\bibitem{cleochic1}
S. Chen {\it et al.} (CLEO Collaboration),
hep-ex/0009044, CLNS 00/1691, CLEO 00-18. 


\bibitem{cleoold}
C.P. Jessop {\it et al.} (CLEO Collaboration),
 Phys. Rev. Lett. {\bf 79} 4533 (1997);
B. Barish  {\it et al.} (CLEO Collaboration),
 Phys. Rev. {\bf D51} 1014 (1995).


\bibitem{cleof00}
J.P. Alexander {\it et al.} (CLEO Collaboration),
CLNS 00/1670, CLEO 00-7





\end{thebibliography}
\end{document}